\documentclass[11pt]{article}

\IfFileExists{srcltx.sty}{\usepackage[active]{srcltx}}

\usepackage{epsfig}
\usepackage{amsfonts}
\usepackage{bbm}
\usepackage{units,slashed}
\usepackage{amsmath,amssymb} %
\usepackage{xspace,bm,paralist,graphicx,color}

\def\be{\begin{equation}}
\def\ee{\end{equation}}
\def\ba{\begin{eqnarray}}
\def\ea{\end{eqnarray}}
\def\lsi{\raise0.3ex\hbox{$<$\kern-0.75em\raise-1.1ex\hbox{$\sim$}}}
\def\gsi{\raise0.3ex\hbox{$>$\kern-0.75em\raise-1.1ex\hbox{$\sim$}}}

\newcommand{\ndecay}{N\to \mu \pi, \mu e \nu}

\newcommand{\C}{1-\cos\Theta} %

\def\l{\left(}
\def\r{\right)}

\begin{document}


\title{
 Search for GeV-scale sterile neutrinos   responsible for 
active neutrino oscillations and baryon  asymmetry of the Universe
}

\author{S.N. Gninenko$^1$, D.S. Gorbunov$^1$, and M.E. Shaposhnikov$^2$}

\date{}
\maketitle

{\em $^1$Institute for Nuclear Research of Russian Academy of Sciences,
117312 Moscow, Russia}

{\em $^2$Institut de Th\'eorie des Ph\'enom\`enes Physiques, EPFL, 
CH-1015 Lausanne, Switzerland}

 

 
\begin{abstract}
Standard model of particle physics fails to explain neutrino
oscillations, dark matter and  baryon asymmetry of the Universe. All
these problems can be solved with three sterile neutrinos added to the
model. Quite remarkably, if sterile neutrino masses are well below the
electroweak scale, this modification --- Neutrino Minimal Standard
Model ($\nu$MSM) --- can be tested experimentally.    We discuss a new
experiment on search for decays of GeV-scale sterile neutrinos, which
in the framework of the $\nu$MSM are responsible for the
matter-antimatter asymmetry generation and for the active neutrino
masses.  If lighter than 2 GeV, these particles can be produced in
decays of charm mesons generated by high energy protons  in a target
of a proton beam-dump experiment, and subsequently decay into light SM
particles. In order to fully explore this sector of the $\nu$MSM, the
new experiment requires data obtained with at least $10^{20}$ incident
protons on target (achievable at CERN SPS in the future), and a big
volume detector constructed from a large amount of identical single
modules, with a total sterile neutrino decay length of few
kilometers.  The preliminary feasibility study for the proposed
experiment shows that it has sensitivity which may either lead to the
discovery of new particles {\em below the Fermi scale} ---
right-handed partners of neutrinos --- or rule out  seesaw sterile
neutrinos with masses below 2 GeV.
 
\end{abstract}


\section{Introduction}
\label{sec:intro}
\begin{figure}[!htb]
\centerline{
\includegraphics[width=0.5\textwidth]
{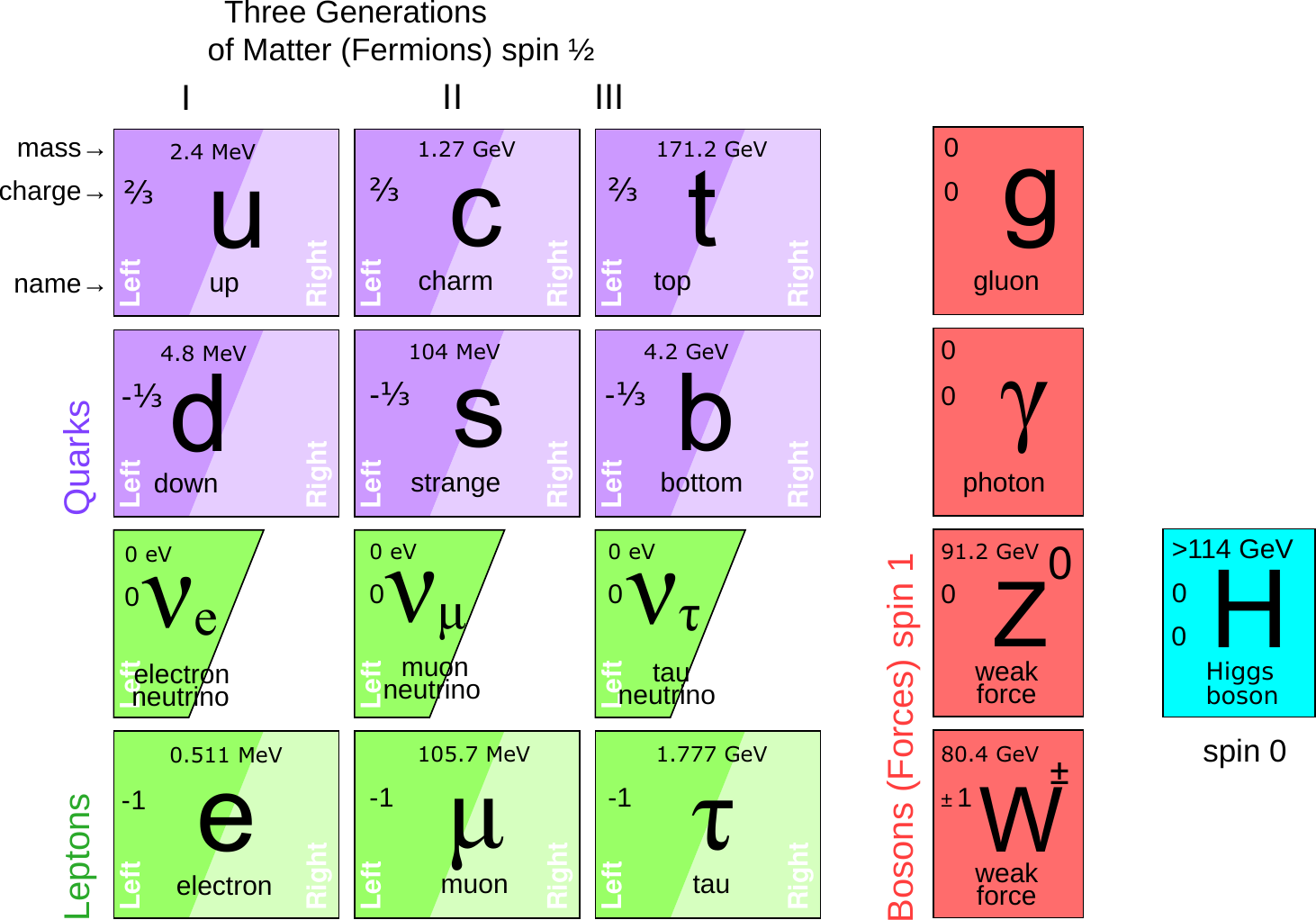}\hspace{0.7cm}
\includegraphics[width=0.5\textwidth]
{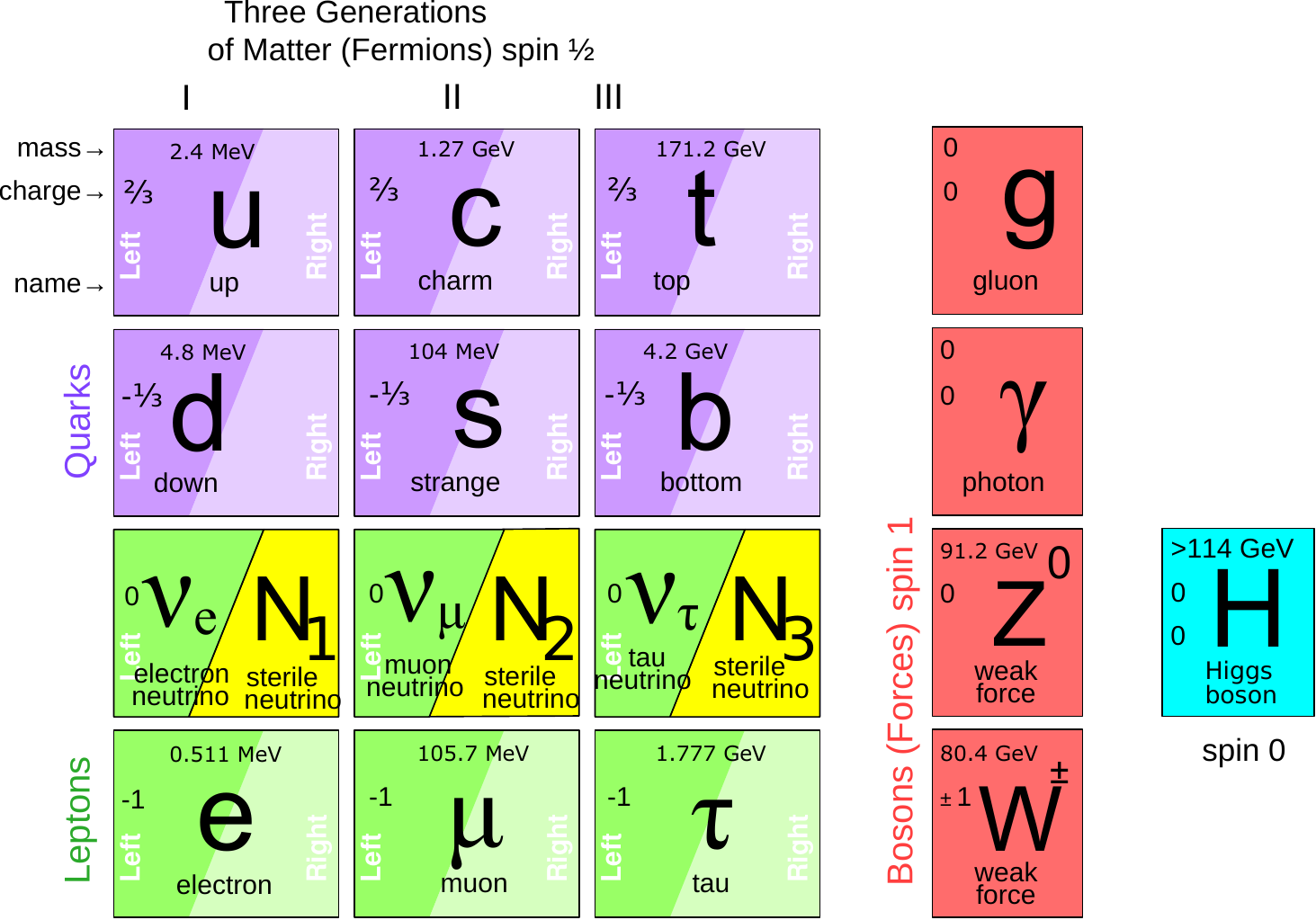}
}
\caption{Particle content of the SM and its minimal extension in neutrino
  sector. In the SM (left) the right-handed partners of neutrinos are
  absent. In the $\nu$MSM (right) all fermions have both left and right-handed
  components.}
\label{ferm}
\end{figure}

The discovery of neutrino oscillations provides an undisputed signal
that the Standard Model (SM) of elementary particles is not complete.
However, what kind of new physics it brings to us remains still
unclear: we do not know yet the properties of new particles which are
believed to be behind this phenomenon. An attractive possibility
is the extension of the SM by three right-handed neutrinos, making the
leptonic sector similar to the quark one, see Fig.\,\ref{ferm}.

\begin{figure}[t]
  \centerline{\includegraphics[width=0.6\textwidth]{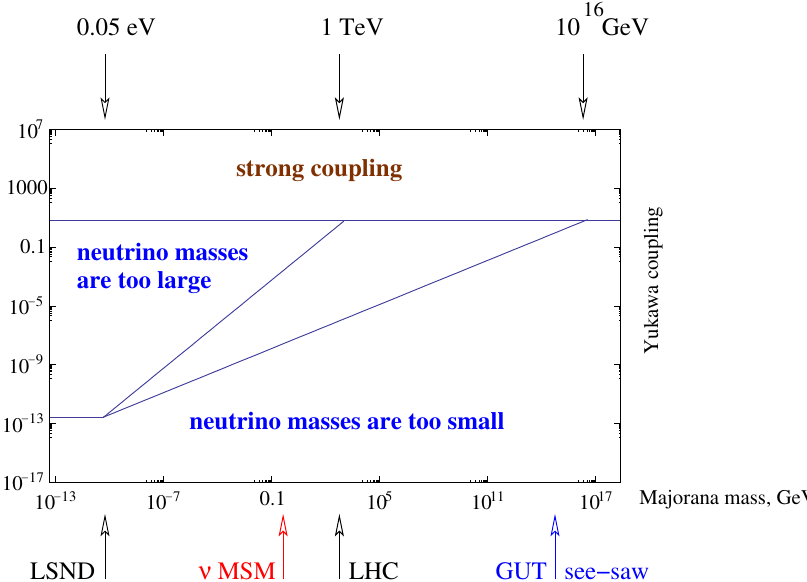} }
\caption{The admitted values of the Yukawa couplings of sterile neutrinos as a
function of their seesaw Majorana masses.
\label{yukawa}
}
\end{figure}

The masses $M$ and Yukawa couplings $Y$ of new leptons remain largely
unknown. Basically, $M$ can have any value between zero (corresponding to
Dirac neutrinos) and $10^{16}$ GeV, whereas $Y$ can vary from $10^{-13}$
(Dirac neutrino case) to $\sim\pi$ (the onset of the strong coupling). The
admitted region is sketched in Fig.~\ref{yukawa}.

Independently on their mass, the new Majorana leptons can explain
oscillations of active neutrinos. So, an extra input is needed to fix
their mass range. It can be provided by the LHC experiments.

Suppose that the resonance found at the LHC by Atlas and CMS in the
region $125-126$ GeV is indeed the Higgs boson of the Standard Model. 
This number is remarkably close to the lower limit on the Higgs mass
coming from the requirement of the absolute stability of the
electroweak vacuum and from Higgs inflation, and to prediction of the
Higgs mass from asymptotic safety of the Standard Model (see detailed
discussion in \cite{Bezrukov:2012sa} and in a proposal submitted to
European High Energy Strategy Group by Bezrukov et al. \cite{BeP}). 
The existence of the Higgs boson with this particular mass tells that
the Standard Model vacuum is stable or metastable with the life-time
exceeding that of the Universe. The SM in this case is a valid
effective field theory up to the Planck scale, and no new physics is
required {\em above} the Fermi scale from this point of view.  Suppose
also that the LHC finds no new particle and no deviations from the
Standard Model. In this case the ``naturalness paradigm'', leading the
theoretical research over the last few decades will be much less
attractive, as the proposals for new physics stabilizing the
electroweak scale by existence of new particles in the TeV region and
based on low energy supersymmetry, technicolor or large extra
dimensions would require severe fine-tunings.

The solution of the hierarchy problem does not require in fact the
presence of new particles or new physics above the Fermi scale.
Moreover, {\em the absence} of new particles between the electroweak
and Planck scales, supplemented by extra symmetries (such as the scale
invariance) may itself be used as an instrument towards a solution of
the problem of stability of the Higgs mass against radiative
corrections (for detailed arguments see 
\cite{Shaposhnikov:2007nj,Shaposhnikov:2008xb,Shaposhnikov:2008xi}).

Even regardless the hierarchy problem, it is clear that the Standard
Model of elementary particles is not complete. It is in conflict with
several observations. These are non-zero neutrino masses and
oscillations discussed above, the excess of matter over antimatter in
the Universe, and the presence of non-baryonic dark matter. Any model
of physics beyond the SM (BSM) should be able to deal with the
experimental troubles of the SM.

\begin{figure}[!t]
\centerline{\includegraphics[width=0.7\textwidth]
{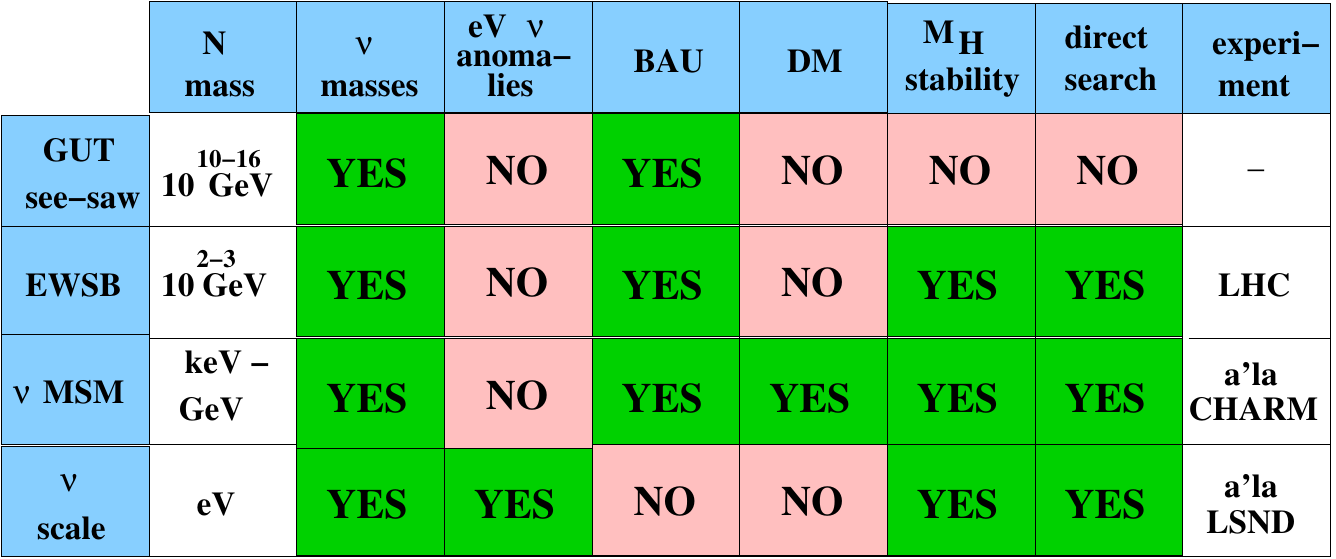}}
\caption{This table shows whether a given choice of the mass of sterile
  neutrinos can explain neutrino masses and oscillations, accommodate eV
  neutrino anomalies, lead to baryogenesis, provide a dark matter candidate,
  ensure the stability of the Higgs mass against radiative corrections, and be
  directly searched at some experiments.  }
\label{table}
\end{figure}

Guided by the arguments steaming from alternative solutions to the
hierarchy problem it is natural to ask whether the observational
problems of the SM can be solved by new physics {\em below} the Fermi
scale.  And the answer is \emph{affirmative}: an economic way to
handle in a unified way the problems of neutrino masses, dark matter
and baryon asymmetry of the Universe is to add to the SM three
Majorana singlet fermions with masses roughly {\em of the order of
masses of known quarks and leptons}. This theory is called the
$\nu$MSM, for ``Neutrino Minimal Standard Model'' (for a review see
\cite{Boyarsky:2009ix}). The lightest of the three new leptons is
expected to have a mass from $1$ keV to $50$ keV and plays the role of
the dark matter particle (see detailed discussion in
\cite{Boyarsky:2009ix} and in a proposal submitted to European High
Energy Strategy Group by Boyarsky et al. \cite{BP}). Two other
neutral fermions are responsible for giving masses to ordinary
neutrinos via the see-saw mechanism at the {\em electroweak scale} and
to creation of the baryon asymmetry of the Universe. The masses of
these particles and their couplings to ordinary leptons are
constrained by particle physics experiments and cosmology. Two leptons
should be almost degenerate, forming thus nearly Dirac fermion (this
is coming from the requirement of successful baryogenesis). For
comparison, we show in Fig.\,\ref{table}
the summary of different possibilities for the masses of Majorana
leptons.

A lot of experimental efforts were devoted to the direct search of
Majorana neutral leptons in the past
\cite{Yamazaki:1984sj}-\cite{Achard:2001qw}.  No new particles were
found, but several constraints on their mixing angles with ordinary
leptons were derived. The interest to these searches declined
considerably at nineties, most probably due to the theoretical
prejudice that the masses of Majorana leptons should be associated
with the {\em Grand Unified scale} $\sim 10^{16}$ GeV
\cite{Seesaw1}-\cite{Seesaw4}, making their direct search impossible. 
As a result, no dedicated searches of relatively light neutral leptons
were performed in the last decade, with an exception of several
experiments \cite{Vaitaitis:1999wq}-\cite{Astier:2001ck} related to
the so-called Karmen anomaly \cite{Armbruster:1995nr}. Today, the
reason for the negative result became clear: in the domain of
parameters accessible to most of these experiments the neutral leptons
interact too strongly to produce the baryon asymmetry of the Universe,
not satisfying one of the Sakharov conditions for baryogenesis.

The recent theory developments provide a guideline for the required
experimental sensitivity \cite{Gorbunov:2007ak}, which was absent at
the time when the early experiments were made.  The aim of the present
work is to discuss the experimental signatures of the existence of
these particles and estimate the parameters of the experiment of
beam-target type required to fully explore the model parameter space
if  sterile neutrinos are lighter than D-mesons  (see a proposal
submitted to European High Energy Strategy Group by Gorbunov and
Shaposhnikov \cite{GS}).  

The paper is organized as follows. In Section\,\ref{sec:model} we give
the essential features of the model and discuss the constraints on new
particles. Section\,\ref{sec:method} is devoted to general description
of experiments on searches for sterile neutrinos of masses in GeV
range.  In Section\,\ref{sec:predictions} we present an estimate of
sensitivity required to test the model and suggest the experimental
setup to accomplish the ultimate task. Section\,\ref{sec:conc}
contains conclusions.


\section{The model and constraints on properties of new particles}
\label{sec:model}
The $\nu$MSM is described by the most general renormalizable
Lagrangian of all the SM particles and 3 singlet fermions. For the
purpose of the present discussion we take away from it the lightest
singlet fermion $N_1$ (the ``dark matter sterile neutrino''), which is
coupled extremely weakly to the ordinary leptons. In addition, we take
$N_2$ and $N_3$ degenerate in mass, $M_2=M_3=M$. The approximate
equality of the masses of $N_{2,3}$ comes from requirement of
resonance production of the baryon asymmetry of the Universe. Then a
convenient parameterization of the mass term and the interaction of
$N$'s with the leptons of SM is given in terms of three parameters
($M$, $\epsilon$, $\eta$):
\begin{equation}
L_{\rm singlet}=\left(\frac{M \sum m_i}{2v^2}\right)^{\frac{1}{2}}
\left[\frac{1}{\sqrt{\epsilon e^{i\eta}}}\bar L_2 N_2 +
\sqrt{\epsilon e^{i\eta}}\bar L_3 N_3\right] \tilde{H}
- M \bar {N_2}^c N_3 + \rm{h.c.} \,,
\label{pmm}
\end{equation}
where $L_2$ and $L_3$ are the combinations of the left-handed doublets
$L_e,~L_\mu$ and $L_\tau$,
\begin{equation}
L_{2}=\sum_\alpha x_{\alpha}L_\alpha~,~~~~
L_{3}=\sum_\alpha y_{\alpha}L_\alpha~,
\label{L23def}
\end{equation}
with $\sum_\alpha |x_{\alpha}|^2=\sum_\alpha |y_{\alpha}|^2=1$. In eq.
(\ref{pmm}) $v=246$ GeV is the vacuum expectation value of the Higgs
field $H$, $\tilde{H}_i=\epsilon_{ij}H_j^*$, $m_i$ are the active
neutrino masses. Note that one of them is negligibly small in
comparison with others in the $\nu$MSM \cite{Boyarsky:2009ix}), so
that $\sum m_i=\kappa m_{atm}$ with $m_{atm}\equiv \sqrt{|\Delta
m^2_{atm}|} \simeq 0.05$ eV being the atmospheric neutrino mass
difference, $\kappa=1~(2)$ for normal (inverted) hierarchy of neutrino
masses.  Mixing parameters $x_{\alpha}$ and $y_{\alpha}$ can be
expressed through the parameters of the active neutrino mixing matrix
(explicit relations can be found in \cite{Shaposhnikov:2008pf}). The
parameter $\epsilon$ (by definition, $\epsilon <1$) and the
CP-breaking phase $\eta$ cannot be determined by neutrino oscillation
experiments.

\begin{figure}[t]
\centerline{
\includegraphics[width=0.45\textwidth]{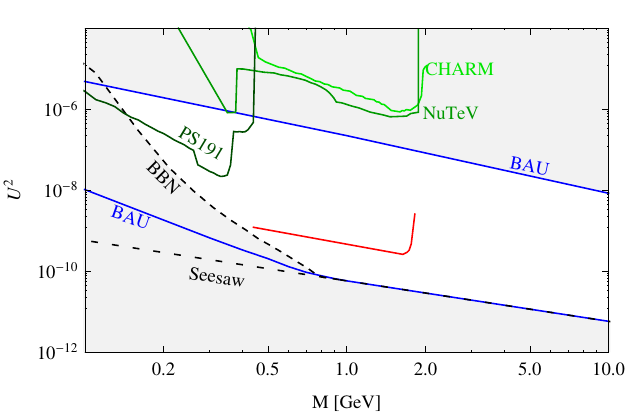}\hskip
0.08\textwidth 
\includegraphics[width=0.45\textwidth]{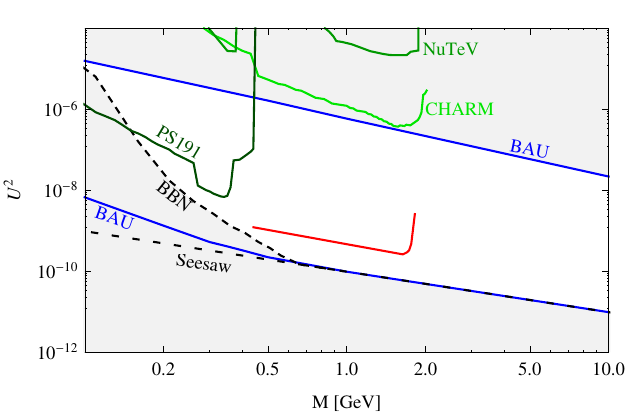}}
\caption{The allowed region of parameters of sterile neutrinos,
  responsible for neutrino oscillations (region above the dotted
  ``see-saw'' line) and for baryo/leptogenesis (region between two
  black solid lines). Sterile neutrinos with the parameters in the
  shaded region to the left of the ``BBN'' line would spoil
  predictions of primordial nucleosynthesis. Accelerator experiments,
  searching for heavy neutral leptons exclude regions above green
  lines. {\it Left panel:} restrictions for normal hierarchy, {\it
  Right panel:} inverted hierarchy. Adopted from Ref.
  \cite{Canetti:2012vf}. The region above the red curve can be probed
  {\bf with a single section of the detector} of length $l_{||}\sim
  100$\,m, height 5\,m and width $l_\bot\sim 5$\,m, placed at a
  distance of about hundred meters; see the text for details.
\label{exp}
}
\end{figure}

\begin{figure}[t]
\centerline{
\includegraphics[width=0.55\textwidth]{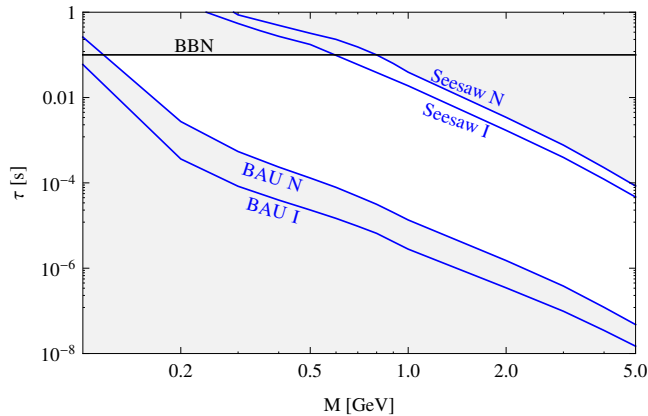}
}
\caption{Constraints on the sterile neutrino lifetime $\tau_N$ coming from the
  baryon asymmetry of the Universe (blue solid lines: ``\emph{BAU N}'' refers
  to the normal hierarchy, ``\emph{BAU I}'' refers to the inverted hierarchy),
  from the seesaw mechanism (blue solid lines: ``\emph{Seesaw N}'' refers to
  normal hierarchy, ``\emph{Seesaw I}'' refers to the inverted hierarchy) and
  big bang nucleosynthesis (black line: ``BBN''). The allowed region of
  parameter space is in white.  Limits from direct searches presented in
  Fig.~\ref{exp} are not outlined.
\label{TauN}
}
\end{figure}

If the value of mass parameter $M$ is fixed, smaller $\epsilon$ yields
stronger interactions of singlet fermions to the SM leptons. This
leads to equilibration of these particles in the early Universe above
the electroweak temperatures, and, therefore, to erasing of the baryon
asymmetry. In other words, the \emph{mixing angle} $U^2$ between
neutral leptons and active neutrinos must be small, explaining why
these new particles have not been seen previously. For small
$\epsilon$, 
\be U^2 = \frac{\sum m_i}{4 M \epsilon}~.
\label{U2}
\ee
The region, where baryogenesis is possible in $(U^2,\,M)$ plane is shown
in Fig.\,\ref{exp}. 
We also plot there the exclusion regions coming  from different
experiments such as  
CHARM \cite{Bergsma:1985is}, NuTeV \cite{Vaitaitis:1999wq}, and CERN PS191
experiment \cite{Bernardi:1985ny,Bernardi:1987ek} (see also discussion of
different experiments in \cite{Atre:2009rg,Ruchayskiy:2011aa}).  Only CERN
PS191 significantly entered into the cosmologically interesting part of the
parameter space of the $\nu$MSM for sterile neutrinos lighter than kaon. 
In Fig. \ref{TauN} 
 we present the expected lifetime of the
singlet fermions in an experimentally interesting region $M<1.8$\,GeV.

As we see, the mass of these particles is limited from below by
current experiments and cosmological considerations by $\sim 100$ MeV,
while no known solid upper bound, better than the electroweak scale,
can be applied. At the same time, various considerations indicate that
their mass may be in  ${\cal O}(1)$ GeV region
\cite{Shaposhnikov:2008pf}. 
\section{General considerations for the experiment}
\label{sec:method}
\begin{figure}[t]
\centerline{\hskip
0.05\textwidth 
\includegraphics[width=0.4\textwidth]{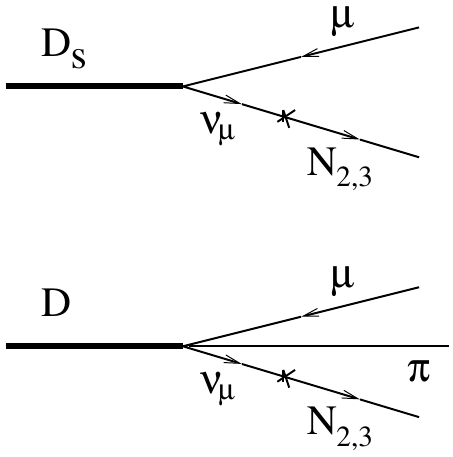}\hskip
0.1\textwidth 
\includegraphics[width=0.4\textwidth]{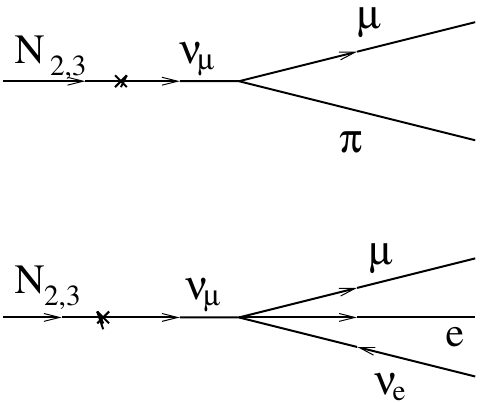}}
\caption{{\it Left panel:} Feynman  diagrams of meson decays to  
heavy sterile neutrinos. {\it Right panel:} Feynman diagrams of sterile
neutrino decays.}
\label{production-and-decays}
\end{figure}

The most efficient mechanism of sterile neutrino production is
associated with weak decays of heavy mesons and baryons, see left
panel of  Fig.~\ref{production-and-decays} 
for examples of relevant two- and three-body decays. Therefore, from
experimental point of view, the distinct mass ranges are associated
with the masses of mesons in which sterile neutrinos can be created:
below $500$ MeV (K meson), between $500$ MeV and $2$ GeV (D-mesons), 
between $2$ and $5$ GeV (B-mesons), and above $5$ GeV. We will
concentrate here on the masses below the charm threshold, the higher
values will be discussed shortly in Section\,\ref{sec:conc}.

The mechanism works as follows.  Heavy mesons (and baryons)  can be
produced by energetic protons scattering off the target material. As
a reference machine we take CERN SPS with beam energy of 400\,GeV. Here
the relevant heavy hadrons contain charm as the heaviest quark: $D$,
$D_s$, $\Lambda_c$, $D^*$, \dots  With typical weak lifetime (in the
rest frame) of about $10^{-10}$\,s these mesons mostly decay before
further interaction in  the target, no matter how thick it is.  
Sterile-active neutrino mixing
gives rise to sterile neutrino production in weak decays of the heavy 
mesons\,\footnote{Note that searches for those heavy hadron decays is 
one of the possible strategy of exploring the $\nu$MSM parameter space 
in full, which at the present level of experimental technique seems
unrealistic, see discussion in Section\,\ref{sec:conc}.}.  
For charmed mesons  typical branching ratios expected
for the $\nu$MSM parameter space  are at the level
of \cite{Gorbunov:2007ak}
\begin{equation}
\label{D-branching}  
\mbox{Br}\l D\to N\r\sim 10^{-8}-10^{-12}
\end{equation}
  (referring to upper and
lower limits in Fig.\,\ref{exp}, respectively).  

\begin{figure}[!htb]
\hskip
0.05\textwidth 
\includegraphics[width=0.9\textwidth]{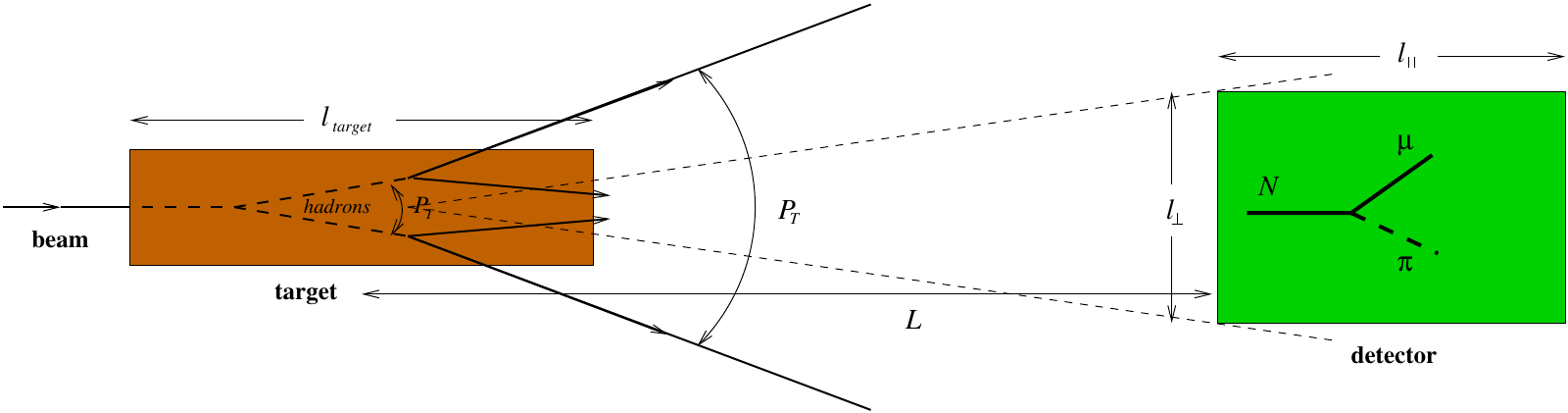}
\caption{Sketch of typical beam-target experiment on searches for
  sterile neutrino decays.}
\label{setup}
\end{figure}

These sterile neutrinos further weakly decay to the SM particles due
to mixing with active neutrinos. In the $\nu$MSM mixing angles are
small, see Fig.\,\ref{exp}, hence the sterile neutrinos live much
longer (see Fig.\,\ref{TauN}) as compared to that of the weakly
decaying SM particles of similar masses. Before the decay (and with
account of $\gamma$-factor) relativistic sterile neutrinos would cover
quite a large distance significantly exceeding ten kilometers.
Hence, sterile neutrino decays into SM particles due to mixing with active
neutrino can be searched for in the near detector, see Fig.\,\ref{setup}.
Two examples of the interesting decay modes are presented on right
panel of Fig.~\ref{production-and-decays}. More examples and typical
patterns of sterile neutrino decay  branching ratios expected within
the $\nu$MSM can be found in  ref.\,\cite{Gorbunov:2007ak}. 


%
Let us first make rough estimates of length-scale of the detector needed to
{\em fully explore} the $\nu$MSM parameter space for sterile neutrino
masses in the interesting range $M_N\sim0.5-2$\,GeV, where sterile
neutrinos are dominantly produced in charmed hadron decays. 
Let $P_T$ and $P_L$ be average transverse and longitudinal momenta of
produced sterile neutrinos. We start with detector of a special
geometry designed to {\em cover entirely} the solid angle $\sim \pi
P_T^2/P_L^2$ around the beam axis. Then all the sterile neutrinos
travel through the fiducial volume of the detector which has a conical
form. For 400\,GeV proton beam operating at SPS the charm production
cross section $\sigma_c$ was measured by LEBC/EHS collaboration
revealing $\sigma_c\approx 30\,\mu$b\,\cite{AguilarBenitez:1987rc}. 
Thus for $N_{\rm pot}=10^{20}$ protons incident on the target per year one
expects 
\[
\frac{\sigma_c}{\sigma_p}\, N_{\rm pot}\sim 0.75\times 10^{17}
\]
charmed hadrons ($\sigma_p\approx 40$\,mb \cite{Nakamura:2010zzi}  is
the total proton cross section at SPS energy). With charm branching
ratio to sterile neutrino given above \eqref{D-branching} one expects
about 
\[
N_N\sim 0.75\times \l 10^5-10^9\r
\]
sterile neutrinos crossing the detector each year of operating. 

Then, for reference sterile neutrino mass of 1\,GeV one estimates the
decay length of $10^2-10^5$\,km (see Fig.\ref{TauN}) and hence 
\[
N_N^{dec}\sim 1-10^7
\]
decay events per year for 1\,km-length detector. Note, that with
$P_T/P_L\sim 1\,\mbox{GeV}/40\, 
\mbox{GeV}$\,\cite{AguilarBenitez:1987rc} 
and heavy  neutrino following the parent meson travel direction, the
detector at a distance of 1\,km has to cover the area of about
$25$\,m$\times25$\,m  to accomplish the task. 

Considering a more realistic setup, where detector is placed on the
Earth surface and its height is of about 5\,m (cf. the size of CHARM
detector \cite{Bergsma:1985is}) one concludes that the detector has to
be about 5 times longer to achieve the same sensitivity to the sterile
neutrinos as in the previous conical configuration.

\begin{figure}[htb!]
\hskip
0.05\textwidth 
\includegraphics[width=0.9\textwidth]{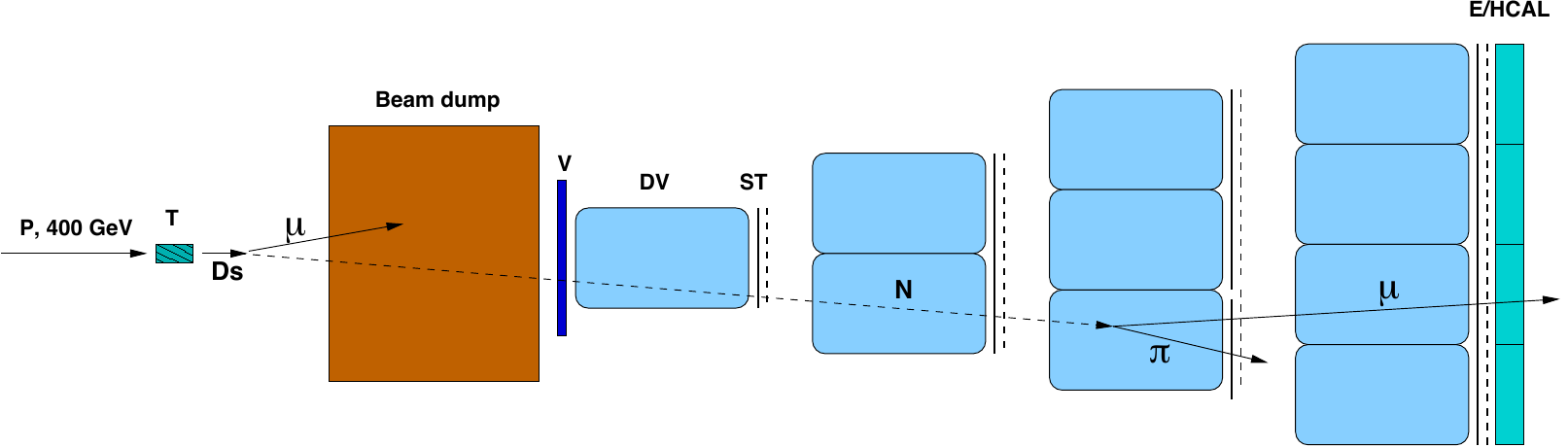}
\caption{Schematic illustration of a proton beam dump experiment on
  search for $D_s \to \mu N, ~šN \to \mu \pi$ decay chain: charm
  mesons $D_s$ generated by the proton beam in the target (T) produce
  a flux of high energy $N$'s through the $U_{\mu N}$ mixing in the
  decay $D_s\to \mu \nu_\mu$, which penetrate the downstream shielding
  and decay into $\mu \pi$ pair in a neutrino decay volume (DV). The
  same setup can be used to search for the process $šN \to \mu e \nu
  $. See text.
\label{setup1}
}
\end{figure}

To register all the expected sterile neutrino decays it seems
appropriate to install many detectors of a reasonable
$5$\,m$\times5$\,m$\times100$\,m size  rather than one large detector.
These small detectors may operate separately by independent groups of
experimentalists but together cover the
same volume as the large detector, see Fig.\,\ref{setup1}.   


\section{Preliminary study for the feasibility of the  proposed experiment }
\label{sec:predictions}

 The searches for Majorana leptons were undertaken in the past at
 PS191, BEBC, CHARM, NuTeV and led to a negative result, not
 surprising in view of the cosmological constraints on the properties
 of sterile neutrinos, that were derived later on.  The decays of the
 $\nu$ MSM sterile neutrinos are very rare events, and their
 observation presents a challenge for the detector design and
 performance. The sensitivity of the search has $|U_{\mu N }|^4$
 dependence on the mixing stength, and in order to increase it
 substantially as compared to the previous searches, one has to use:
\begin{itemize}
\item higher intensity of the proton beam and larger amounts of pot's.
\item shorter distance to the proton target 
\item larger $N$ decay fiducial volume 
\item longer time of data accumulation
\item better efficiency of events detection, reconstruction, and background rejection, etc... 
\end{itemize}
To compromise the cost of the detector and its performance, we focus
mainly on discussions of the experimental setup to search for the $N$
decays in a detector volume filled with a (diluted) gas, not in  vacuum. The  setup with vacuum requirements would provide possibly  better sensitivity, but is more complicated and costly.

The main components of the experimental setup to search  for the $N$ 
decays of sterile neutrino  are schematically illustrated in Fig. \ref{setup1}.
The full detector is composed of single identical modules. Each module consists of a decay volume tank  (DV)
with diameter of about 5-7 meters, and length of 20-30 meters.  
To minimize cost and amount of passive material, the decay tank could be, e.g. a rubber ballon filled with helium. 
Each DV is followed by a high precision tracker, which could be a layers of straw tube (ST),
 see e.g. \cite{petti}.  
The entire detector is assembled from $\simeq$ 50-100 such modules and is followed by a
calorimeter (E/HCAL) which has  the electromagnetic and hadronic parts. The detector is protected against 
charged secondaries  from neutrino interactions in the dump by a Veto (V).

The design of a single module is similar to that one used in the  PS191, or CHARM experiments, that 
searched for  decays $N \to \nu e^+ e^-, \mu \pi, \mu e \nu, ..$ of heavy neutrinos
in the $N$ mass range from 10 MeV to 1.8 GeV originated from  decays $\pi, K $ and charmed $D$ mesons decays
\cite{Bergsma:1985is, Bernardi:1985ny, Bernardi:1987ek}. The later experiment, specifically 
designed to search for neutrino decays in a high-energy neutrino beam,  was performed 
by   using 400 GeV protons 
from the CERN Super Proton Synchrotron (SPS) with the  total number of $2.4\times 10^{18}$ 
protons on (Cu) target (pot). 
The CHARM decay detector (DD), located at the distance of 480 m from the target,
 consists of decay volume of $3\times 3\times 35$ m$^3$ , three  chambers modules located 
inside the volume to detect charged tracks and  followed by a calorimeter.
The decay volume  was essentially  an  empty region to substantially reduced the number of ordinary neutrino interactions. 
The signature of the heavy neutrino decay $\nu_h \to \nu e^+ e^- $ were events originating in the decay region at a small angle with respect to the neutrino beam axis with one or two separate electromagnetic showers in the calorimeter 
\cite{Bergsma:1985is}.
No such events were observed and limits were established on the 
$\nu_{e,\mu} - \nu_h$ mixing strength as a function of the $\nu_h$ mass.

\subsection{Production and decay of sterile neutrinos}
If the  decays $D_s, D \to \mu  N$ exist, one expects a  flux of 
high energy $N$'s from the SPS target, since neutral mesons $D_s$ and $D$ are
abundantly produced in the forward direction  by high energy  protons in the target.
If $N$ is a relatively long-lived particle, this flux  would penetrate the downstream 
shielding without significant attenuation  and would be observed in the proposed  
detector via the $\ndecay$ decay into a high energy $\mu \pi$ or $\mu e $ pair, as schematically 
illustrated in Fig. \ref{setup}.
The experimental signature of the sterile
neutrino decays is an observation of either $\mu \pi$ or $\mu e $
pairs originated from a common vertex located in empty space. This signature is clean and the signal events are expected 
to be selected in the detector  with a small background. 
The occurrence of $\ndecay$ decays  would appear as an excess
of $\mu \pi $ or $\mu e$ pairs in the detector  above those expected from  standard neutrino interactions. 
Up to a $N$ mass of $\lesssim$ 2 GeV , the mass difference between the $D_s$ 
and the $\mu$, $N$'s can originate both from the $\nu_{\mu}$
produced directly in the $D_s$ decay and from the $\nu_{\mu}$ 
produced indirectly in the subsequent decay of  $\tau$ produced in $D_s$ decays.\ For a $N$ mass 
below  than $\lesssim$ 1.8 GeV $\nu_{\mu}$'s produced in  
$\tau$ decay can contribute.\ However these indirect $N$'s 
have a lower acceptance  resulting in a 
smaller probability to observe the $N$ decays in the  
detector.\ Therefore we focus mainly on directly produced $N$'s. 

As discussed previously, the  mixing between the sterile neutrino and the muon neutrino 
results in the  decay $D_s \to \mu N$, as illustrated  in Fig.~\ref{setup1}. 
The $D_s$ meson, which is normally decays into a 
$\mu$  and $\nu_\mu$, might instead decay to a $\mu$ and a 
heavy neutrino $N$. For the  mass interval $m_{\nu_h}\lesssim 2$~GeV 
the chirality-flip is mostly due to sterile neutrino mass  which results in 
\begin{equation}
\Gamma(D_s \to \mu N) \approx 
\Gamma(D_s \to \mu^ \nu_\mu)|U_{\mu N}|^2
\Bigl(\frac{m_{N}}{m_\mu}\Bigr)^2\;.
\label{rate3}
\end{equation} 
In the SM, $D_s$ meson decays leptonically via annihilation of the $c$
and $\overline{s}$ quarks through a virtual $W^+$.  The decay rate of
this process is given by
\begin{equation}
\Gamma(D_s \to l \nu) = \frac{G_F^2}{8 \pi}f^2_{D_s} m_l^2
M_{D_s} \!\!
\l \!\!1-\frac{m_l^2}{M_{D_s}^2} \!\r^{\!\!2} \!\!|V_{cs}|^2\;,
\label{rate1}
\end{equation}
where the $M_{D_s}$ is the $D_s$ meson mass, 
$m_l$ is the mass of the charged 
lepton, $f^2_{D_s}$ is the decay constant, $G_F$ is the Fermi 
constant, and $V_{cs}$ is a Cabibbo-Kobayashi-Maskawa matrix element 
which value equals 0.97334 \cite{pdg}. The decay rate \eqref{rate1} 
is suppressed by the lepton mass squared, since the
very leptonic decay is due to chirality-flip. 
As follows from Eq.(\ref{rate3}), in neutrino scattering experiments the $N$  decay 
signal rate is proportional to  $\propto |U_{\mu N}|^4$  (the mixing  $|U_{\mu h}|^2$ 
appears twice,  through the heavy neutrino production,  and  
 through the its decays in the detector)
and, thus is more suppressed. The branching ratio of a particular decay mode, e.g. $N\to \mu \pi$ (or $N\to \mu e \nu$) is given by  
\begin{equation}
BR(N \rightarrow \mu \pi) = 
\frac{\Gamma(\mu \pi)}{\Gamma_{tot}} = F(M_N,|U_{\mu N}|^2)\;,
\end{equation}
where the function $F(M_N,|U_{\mu N}|^2)$ is calculated in \cite{Gorbunov:2007ak}, and the total rate $\Gamma_{tot}$ is dominated by  the $N \to 3 \nu  $ decay channel.

To make quantitative estimates, we performed simplified simulations of
the $N$ production in the inclusive reaction
\begin{equation}
p + Be \to D_s + X, ~ D_s \to \mu + N \to \mu \pi , \mu e \nu
\label{reaction}
\end{equation}
with the emission of a GeV-scale sterile neutrino $N$ subsequently
decaying semi-leptonically or leptonically in the detector, as illustrated in Figure
\ref{setup},\ref{setup1}. In the simulations it
is assumed, that the $N$ is a long-lived particle and the flux of $N$'s
penetrates the downstream shielding without significant
attenuation.  The decay $\ndecay$
cannot be distinguished from the anti-neutrino decay $\overline{N}\to
\mu \pi$ or $\overline{N}\to \mu e \nu $ and the obtained result 
therefore refers to the sum of all these decay modes.

The flux of sterile neutrinos from decays of $D_s$'s and $D$'s
produced in the target by primary protons can be expressed as follows:
\begin{equation}
\Phi(N)\propto N_{\rm pot}\int \frac{d^3\sigma(p + N\to D_s (D) + X)}{d^3p_{D_s(D)}} 
\, {\rm Br}(D_s (D)\to \mu N)\,f\, d^3p_{D_s(D)}  \;,
\label{flux}
\end{equation}
where $N_{\rm pot}$ is the  number of pot, $\sigma(p + N\rightarrow D_s (D) + X)$
is the $D_s (D)$ meson production cross-section,
${\rm Br}(D_s (D) \to \mu N)$ is 
 the $D_s (D) \to \mu N $ decay mode branching fraction \cite{pdg}, and
$f$ is the decay phase space factor, respectively.

To perform quantitative  we used simulations of the process shown in Fig.\ref{setup} by taken into account  
the relative normalization of the yield of charmed meson species $D_s$ and $D$ from the original 
publications. 
The invariant cross section of a charm meson   production  can be expressed as \cite{pdg}
\begin{equation}
E\frac{d^3 \sigma }{d^3p} = 
\frac{d^3 \sigma }{p_Tdp_T dy d\phi} = \frac{d^2 \sigma }{2\pi p_T
dp_T dy}\;,
\label{crsec}
\end{equation}
where $p_T$ is the transverse momentum of the particle, $y$ is its rapidity, and 
 in the last equality integration over the full 2$\pi$
azimuthal angle $\phi$ is performed. For the production cross sections of $D_s$ and $D$ charm mesons we used the Bourquin-Gaillard (B-G) formula from Ref.\cite{bourq}:
\begin{equation}
E\frac{d^3\sigma(p + N\rightarrow D_s (D) + X)}{d^3 p}= A \bigl(\frac{2}{E_T+2}\bigr)^{12.3} {\rm exp}(-\frac{5.13}{Y^{0.38}}) f(p_T), 
\label{bourq}
\end{equation}
where 
\[f(p_T) = \left\{ 
\begin{array}{l l}
  exp(-p_T) & \quad \mbox{ $p_T<1$ GeV}\\
  exp(-1-23(p_T-1)/\sqrt{s}) & \quad \mbox{ $p_T>1$ GeV}\\ \end{array} \right. \]
with $Y=y_{max}-y$, being the rapidity, $s$ is the Mandelstam variable in GeV$^2$, and $E_T(p_T)$ is  the transverse energy (momentum) in GeV.
The coefficient $A$ in \eqref{bourq} is a normalization factor that was
tuned to obtain the  cross sections of the charm production in pp collisions 
at 400 GeV. The spectra of $\nu_{\tau}$'s produced in the Be target
by primary protons could be calculated by using the 
approach reported in ref. \cite{concha} (see also \cite{van}). 
However, the Bourquin-Gaillard (B-G) approach gives the parametric form of 
\eqref{crsec} for the production of many different hadrons, including charm,  which is in good agreement with data
over the full phase-space in high-energy hadronic collisions and with a precision which is satisfactory at the level of  
our accuracy.
The total $D_s$ and $D$  production cross sections in p-Be collisions were calculated  from its linear extrapolation to the target atomic number. The contribution of protons not interacting 
in the Be target and interacting in the SPS beam dump was not taken into account.

\begin{figure}[t]
\hskip
0.05\textwidth 
\includegraphics[width=0.9\textwidth]{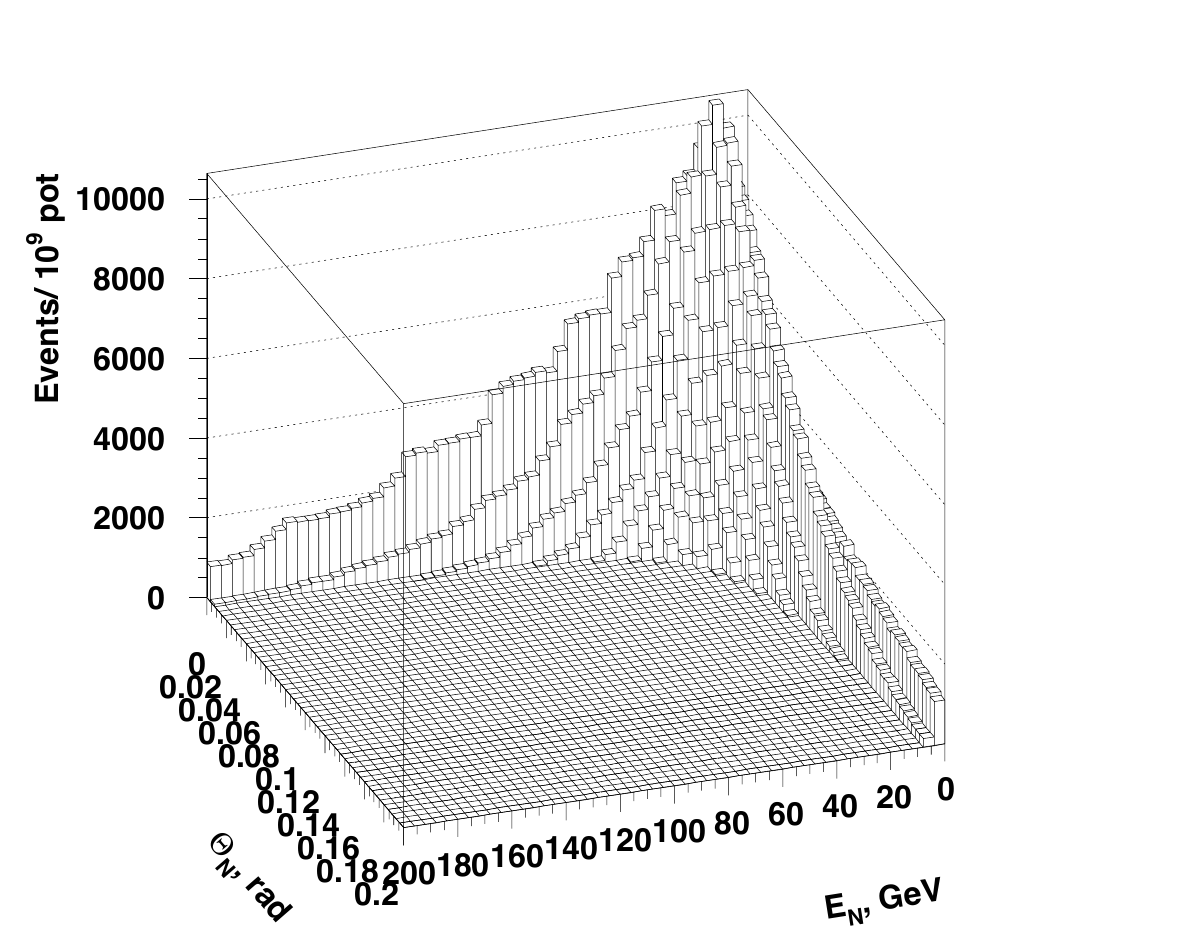}
\caption{Two dimensional distribution of flux for 1 GeV sterile neutrino  as a function of 
$\Theta_N$, the angle between the 
proton beam axis and momentum of $N$, and its energy calculated for 400 GeV proton incident on Be target.
The N's are produced through the decay $D_s \to \mu N$ assuming mixing strength $|U_{\mu N}|^2 = 1$.  
\label{bipang}
}
\end{figure}

The calculated fluxes and energy distributions  of $D_s, D$ produced in the Be target were used to predict the  
flux of $N$'s, as a  function of its mass. An example of 2D-flux distribution calculated 
for 1 GeV sterile neutrino  as a function of  $\Theta_N$, the angle between the 
proton beam axis and momentum of $N$, and the N energy  for 400 GeV proton incident on target is shown in 
Fig. \ref{bipang}. In these calculation the 
 N's are produced through the decay $D_s \to \mu N$ assuming mixing strength $|U_{\mu N}|^2 = 1$.
 One can see, that for detector located at the distance $\simeq$ 100 m from the target its fiducial area 
 for efficient $N$ detection should be of the order $S \simeq \pi L^2 \Theta^2 \simeq$300 m$^2$.

\subsection{Background}

 There are several expected sources of background in the experiment, which come from 
 \begin{itemize}
 \item neutrino interactions in hellium
 \item neutrino interactions in surrounding materials 
 \item cosmics
 \item accidentals
 \item punch-through energetic neutral particles, such as $K_L^0$ or $n$
 \end{itemize}
Because of the high number of pot's and many possible sources of background  
the detail study for the feasibility of the proposed search   would require the simulation of 
a very large number of events resulting in a prohibitively large amount of 
computer time.
Consequently, only small fraction  of the required statistics 
for $\nu_{\mu}CC(NC)$ inelastic reactions and other
background components, e.g. such as coherent and  
quasi-elastic reactions, etc...,  
were considered for preliminary estimate and combined with numerical calculations.  

 The estimate shows that the dominant background is expected from the primary $\nu_\mu$ interactions 
 in passive material, which correspond roughly to $\simeq 10^6~\nu_\mu CC$/$10^{20}$ pot  events per 1 ton of matter.
 The  largest contribution to background is expected  from
 neutrino interactions yielding a single 
$\mu \pi$ pair  with little hadronic activity in the final state.\
 Neutrino interactions that occur  in the surrounding material, e.g. upstream 
 of the DV fiducial volume or in material of the decay tanks, also  may produce an isolated $\mu \pi$ pairs if the  
accompanying particles are not detected.
These   background events  from surrounding material could  be rejected with 
an  effective method developed in Ref.\cite{Astier:2001ck}, by introducing the collinearity variable $\C$, where 
$\Theta $ is the angle between, e.g. the $\mu \pi$ pair momentum direction and the neutrino beam axis. 
The main idea is that if the $\mu \pi $ pair originates from the decay chain $D_s \to \mu N, ~N\to \mu \pi$,  at high energies the $\mu \pi$ pair momentum extrapolated back should point 
to the $D_s$ production target. An example of the distribution of the variable $\C$ obtained in the NOMAD experiment 
searching for the decay $N\to e^+ e^- \nu $,  is shown in Figure 5 of Ref.\cite{Astier:2001ck}. One can see that 
potentially one could measure the $\mu \pi$ pair momentum direction with precision  $\C \simeq 10^{-5}$.
By using this variable, two independent techniques could be  used  for the 
background estimation in the signal region.  The first is obviously based on the MC.
The second method relies on the data themselves.
By extrapolating the  observed events to the signal region with 
the shape of the fully simulated MC events a  
background estimate can be obtained from the data. Using this technique 
would results in an additional  rejection of background events.

\subsection{Sensitivity estimate} 

For a given flux $d\Phi(M_{N}, E_{N}, N_{\rm pot})/d E_{N}$ of $N$'s the expected number of 
$\ndecay$ decays occurring within the fiducial length $L$ of the 
detector located  at a distance $L'$ from the
neutrino target is given by 
\begin{eqnarray}
N_{\ndecay}=Br (D_s \to \mu N)  Br(N \to \mu \pi ) \int \frac{d\Phi}{dE_{N}}\nonumber \\  
\cdot  {\rm exp}\Bigl(-\frac{L'M_{N}}{P_{N}\tau_{N}}\Bigr)
\Bigl[1-{\rm exp}\Bigl(-\frac{L M_{N}}{P_{N}\tau_{N}}\Bigr)\Bigr]\zeta 
S dE_{N} \;,
\label{nev}
\end{eqnarray}
where $ E_{N}, P_{N}$, and $\tau_{N}$ are the $N$ energy, momentum and the lifetime  
at rest, respectively, and  $\zeta$($\simeq 50$\%) is the $\mu \pi$ ($\mu e$) pairs reconstruction efficiency.
The acceptance $S$ was calculated by tracing $N$'s
produced in the Be-target to the detector taking the
relevant momentum and angular distributions into account. 
In this estimate the  average momentum is $<p_{N}>\simeq 45$ GeV , and $S\simeq 50$\%.

\begin{figure}[t]
\hskip
0.05\textwidth 
\includegraphics[width=0.7\textwidth]{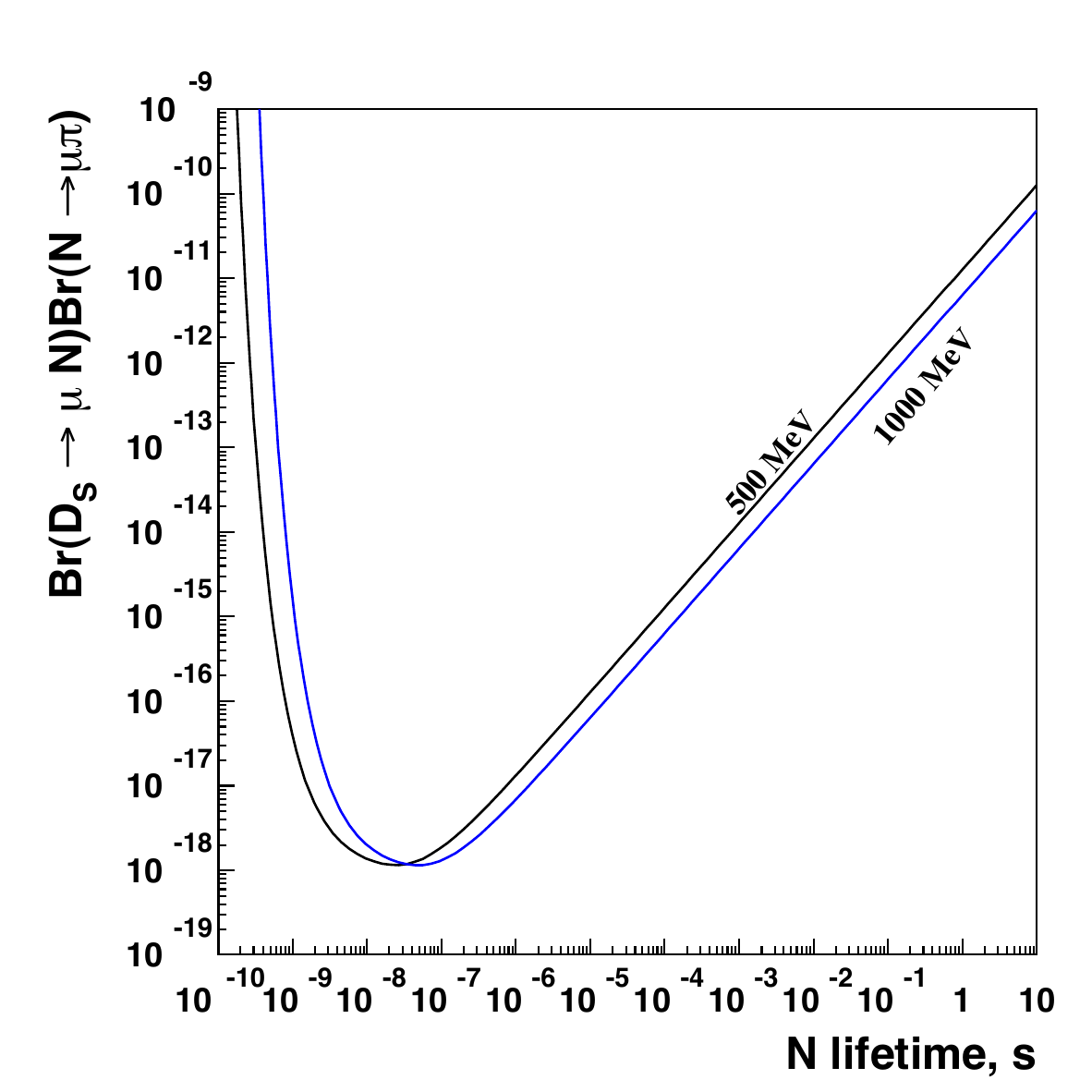}
\caption{ The expected 90$\%~C.L. $  upper limits on the branching ratio
$Br(D_s \to \mu N) Br(N\to \mu \pi)$ versus $\tau_{N}$ estimated  for  the proposed experiment in the background free case. The numbers  near the curves indicate the corresponding values of sterile neutrino masses.  
\label{bratio}
}
\end{figure}
In the case of no signal observation, the background free experiment would result in stringent constraints 
on existence of the sub-GeV sector of the $\nu$MSM. An example of the exclusion region calculated   
in the ($Br(D_s \to \mu N) Br(N\to \mu \pi); \tau_N$) plane for $\simeq 10^{20}$ pot and the detector assembled from $\simeq 10^2$ modules, assuming  $N_{\ndecay} < 2.3$ at 90\% CL,   is  shown in Fig. \ref{bratio}.
For the mass range $ 0.5 \lesssim m_N \lesssim 2 $ GeV  these limits  corresponds to the bounds 
on tne mixing strength which are in the range $ 10^{-6} \lesssim
|U_{\mu N}|^2  \lesssim 10^{-10} $ and exclude seesaw sterile neutrinos
with masses below 2 GeV.

Note, that the  uncertainty for the $D_s$ production at low $p_T$ does not significantly affect the sensitivity estimate. For example, the variation of the $D_s$ yield in \eqref{nev}
by 30\% results in the corresponding variation of the limits on mixing strength $U_{\mu N}$  of order 5\%. 
This is because the sensitivity of the search is proportional to the $|U_{\mu N}|^4$. Indeed, in \eqref{nev} the 
branching fraction of \eqref{rate1} and  the decay  rate $\Gamma (\ndecay)$
of \eqref{rate3} both are  proportional to  $|U_{\mu N}|^2$.

\section{Conclusions} 
\label{sec:conc} 
The energy scale of new physics is not known. If it exists at energies
above the Fermi scale (examples include supersymmetry, large or warped
extra dimensions, models with dynamical electroweak symmetry
breaking), the  search for new particles can be carried out in direct
experiments, such as ATLAS and CMS at LHC. In addition, new
hypothetical heavy particles inevitably appear as virtual states,
leading to different rare processes, absent in the SM. These effects
can be found at experiments such as LHCb and are competitive with the
direct searches. If no new physics is found at the LHC, this may
indicate the absence of new physics all the way between the Fermi and
Planck scales. 

Quite paradoxically, the largely unexplored up to now domain of
energies where the new physics can be hidden is related to physics {\em
below} the Fermi scale. If the new particles are light and weakly
interacting (as in the $\nu$MSM, discussed here) then the search for
rare processes is superior to high energy experiments. It provides a
unique possibility for discovery of new physics, not accessible by any
of the LHC experiments. 

The present proposal deals with the searches of new relatively light
Majorana leptons which "give" the masses to ordinary neutrinos and
produce baryon asymmetry of the Universe. We discussed here the
parameters of the beam-target experiment which can explore completely 
the mass range below $2$ GeV. Due to the weak coupling of these
particles to other fermions these experiments are very challenging and
require construction of a kilometer scale detector. However, an
experiment at much smaller scales, with 100 meters detectors, would be
able to explore the dominant fraction of the parameter space.

Below we describe shortly the prospect for other experiments and
higher range of sterile neutrino masses. 

We start with remark that signal statistics in beam-target experiment
proposed in this paper is determined by the {\it fourth power of
sterile-active neutrino mixing,} $U^2\times U^2$, since both sterile
neutrino production in meson decays and rate of subsequent neutrino
decay within a detector are proportional to $U^2$. Experiments on
searches of meson decays to sterile neutrinos requires (potentially)
less  statistics of mesons, since the signal statistics are
proportional only to the {\it second power} of sterile-active neutrino
mixing $U^2$.  

With expected for $\nu$MSM branching ratios\,\eqref{D-branching} the
relevant experiment has to collect not less than $10^{12}$ hadrons in
order to fully explore the entire part of parameter space at
$M_N<2$\,GeV. This statistics is far above what has been obtained at
previous $c$- and $b$-factories (CLEO(-c), BaBar, Belle) or at
proposed future $c$-$\tau$ (Novosibirsk) and $B$-factories (Tsukuba,
Frascati), where total number of $D$-meson is planned at the level of
$10^9$, see e.g.\cite{Meadows:2011bk}.  
%
%
%
%
Only at LHC the $D$-meson production rate is enough to perform this
search, however, limits from data taking rate, QCD-background, vertex
resolution, particle misidentification etc, prevent measurement of
$D$-meson branching ratios at the required
level\,\eqref{D-branching}. 

As we have already mentioned, there are no solid-motivated upper
limits on sterile neutrino mass of sub-electroweak scale. Sterile
neutrinos can be heavier than $D$-meson, and then charmed hadrons are
out of use. If sterile neutrinos are lighter than 5\,GeV, they can be
produced in beauty hadron decays. This mass range is 
favorable\,\cite{Canetti:2012vf} for a  particular mechanism of dark
matter production (resonant oscillations in the primordial plasma with
lepton asymmetry) in the early Universe. There are other options for
dark matter production, which are insensitive to the 1\,GeV scale
sterile neutrinos,   see e.g.
\cite{Shaposhnikov:2006xi,Bezrukov:2009yw}.  Typical branching ratios
of $B$-mesons to sterile neutrino expected within $\nu$MSM are
\cite{Gorbunov:2007ak}  by one-two orders of magnitude smaller, than
those of  $D$-mesons, as both upper and lower bounds on mixing $U^2$
decrease with increase of sterile neutrino masses, see
Fig.\,\ref{exp}.    Proposed upgrade of LHCb \cite{LHCb:upgrade}  will
be able of entering the interesting region of $\nu$MSM parameter space
by searching for $b$-meson decays into sterile neutrinos, but can not
cover the entire region. Once can further discuss an idea of using
LHC-beam (4\,TeV, 7\,TeV, higher) for a beam-target experiment of the
kind proposed in this paper, in order to fully explore the $\nu$MSM
parameter space, if sterile neutrino mass belongs to the the interval
2-5\,GeV. 

The preliminary analysis  shows that the
quoted  sensitivity could be obtained with a proton beam and a detector  optimized for  
several its properties. Namely, i) the primary beam intensity and the total number of pot's 
 ii) the fiducial decay volume  of the detector, iii)
the composition of the material of decay tank and  surrounding material and  iv) the
efficiency and precision of  the tracking system,  v) the E/HCAL granularity, energy resolution
and particle identification, and vi) different sources of background  a are of importance.
  
  As far as the cost concerned, each module  may be relatively cheap, empty-space with simple
tracker system either inside or behind the decay volume, and the calorimeter at the far end. Its design may
repeat the design of the  experiment on precision neutrino measurements and searches for sterile
neutrino decays at FNAL  \cite{petti}. The detector
of the 100 length has a sensitivity to $U^2\times U^2$ worth 
by a factor only $50$. Thus even one (the first) section 
will be able to deepen considerably into the
parameter range interesting for cosmology, see Figs. \ref{exp}.

The work was supported in part by 
the grant of the President of the Russian Federation NS-5590.2012.2,
by the SCOPES program (D.G. and M.S.) and 
by the RFBR grant 11-02-01528a (DG). 



\end{document}